\documentclass[a4paper]{jpconf}
\usepackage{graphicx}
\begin{document}
\title{Parton-hadron  matter in- and out-off equilibrium}

\author{E L Bratkovskaya$^{1,2}$, V Ozvenchuk$^2$,  W Cassing$^2$, V P Konchakovski$^2$  and O Linnyk$^2$}

\address{$^1$ Institute for Theoretical Physics, Frankfurt University,
          Frankfurt-am-Main, Germany }
\address{$^2$ Frankfurt Institut for Advanced Studies,
         Frankfurt University, Frankfurt-am-Main, Germany}
\address{$^2$ Institute for Theoretical Physics, University of Giessen, Giessen, Germany}

\ead{Elena.Bratkovkaya@th-physik.uni-frankfurt.de}

\begin{abstract}
We study the shear and bulk viscosities of partonic and hadronic matter - as well as the electric conductivity -
as functions of temperature $T$
within the Parton-Hadron-String Dynamics (PHSD) off-shell transport approach. Dynamical hadronic and
partonic systems in equilibrium are studied by the PHSD simulations in a finite box with periodic boundary
conditions. The ratio of the shear viscosity to entropy density $\eta(T)/s(T)$ from PHSD shows a minimum
(with a value of about 0.1) close to the critical temperature $T_c$, while it approaches the perturbative
QCD (pQCD) limit at higher temperatures in line with lattice QCD results. For $T<T_c$, i.e. in the hadronic
phase, the ratio $\eta/s$ rises fast with decreasing temperature due to a lower interaction rate of the
hadronic system and a significantly smaller number of degrees-of-freedom. The bulk
viscosity $\zeta(T)$ -- evaluated in the relaxation time approach -- is found to strongly depend on the
effects of mean fields (or potentials) in the partonic phase. We find a significant rise of the ratio
$\zeta(T)/s(T)$ in the vicinity of the critical temperature $T_c$, when consistently including the
scalar mean-field from PHSD, which is also in agreement with that from lQCD calculations. Furthermore,
we present the results for the ratio $(\eta+3\zeta/4)/s$, which is found to depend non-trivially on
temperature and to generally agree with the lQCD calculations as well. Within the PHSD calculations,
the strong maximum of $\zeta(T)/\eta(T)$ close to $T_c$ has to be attributed to mean-fields (or potential)
effects that in PHSD are encoded in the temperature dependence of the quasiparticle masses, which is related
to the infrared enhancement of the resummed (effective) coupling
$g(T)$. We also find that the dimensionless ratio of the electric conductivity over temperature
$\sigma_0/T$ rises above $T_c$
approximately linearly with $T$ up to $T=2.5 T_c$, but approaches a
constant above $5 T_c$, as expected qualitatively from perturbative QCD (pQCD).
\end{abstract}

\section{Introduction}
High energy heavy-ion reactions are studied experimentally and
theoretically to obtain information about the properties of nuclear
matter under the extreme conditions of high density and/or temperature.
Ultra-relativistic heavy ion collisions at the BNL Relativistic
Heavy Ion Collider (RHIC) and the Large Hadron Collider (LHC) at
CERN  have produced a new state of matter, the  quark
gluon plasma (QGP). The produced quark gluon plasma behaves as a
strongly-interacting fluid unlike a weakly-interacting
gas \cite{StrCoupled1,StrCoupled2}. Large values of the
azimuthal asymmetry of charged particles in momentum space, in particular the elliptic flow
$v_2$, observed in these experiments
\cite{STAR,PHENIX,BRAHMS,PHOBOS,ALICE} could quantitatively be well
described by means of ideal hydrodynamics up to transverse momenta
on the order of 1.5 GeV/c
\cite{IdealHydro1,IdealHydro2,IdealHydro3,IdealHydro4,IdealHydro5}.
An ideal fluid has been defined as having a zero shear
viscosity $\eta$; yet semiclassical arguments have been given suggesting
that the shear viscosity cannot be arbitrarily small
\cite{NonzeroViscosity}. Indeed, the lower bound for the shear
viscosity to entropy density ratio $\eta/s \geq 1/4\pi$ was
obtained by Kovtun-Son-Starinets (KSS) \cite{KSS} for infinitely
coupled supersymmetric Yang-Mills gauge theory based on the AdS/CFT
duality conjecture. Recent relativistic viscous hydrodynamic
calculations using the Israel-Stewart framework require a very small
$\eta/s$ of $0.08-0.24$ in order to reproduce the RHIC elliptic flow
$v_2$ data
\cite{ViscousHydro1,ViscousHydro2,ViscousHydro3,ViscousHydro4}. The
main uncertainty in these estimates results from the equation of
state and the initial conditions employed.

There is strong evidence from atomic and molecular systems that
$\eta/s$ should  have a minimum in the vicinity of the phase transition
(or rapid crossover) between the hadronic matter and the quark-gluon plasma
\cite{Minshear}, and that the ratio of bulk viscosity to entropy
density $\zeta/s$ should be maximum or even diverge at a
second-order phase transition
\cite{MaxBulk1,MaxBulk2,MaxBulk3,MaxBulk4,MaxBulk5}.

The shear and bulk viscosities of strongly-interacting systems have
been calculated within different approaches. Calculations have been
performed at extremely high temperatures, where perturbation theory can
be applied \cite{pQCD1,pQCD2}, as well as at extremely low
temperatures \cite{pQCD2,pQCD3,pQCD4}. First results for shear and
bulk viscosities obtained within lattice QCD simulations just above
the critical temperature of pure gluon/glueball matter have been published in
\cite{lQCDtransport1,lQCDtransport2,lQCDtransport3,lQCDtransport4}.
In the literature there are several methods for the calculation of the
shear and bulk viscosities, such as the Relaxation Time
Approximation (RTA) \cite{RTA}, the Chapmann-Enskog (CE) method
\cite{CE} and the Green-Kubo method \cite{Green,Kubo}.

In this contribution we extract the shear and bulk viscosities from the
`infinite' parton-hadron matter employing different methods within
the Parton-Hadron-String Dynamics (PHSD) transport approach
\cite{PHSD1,PHSD2}, which is based on generalized transport
equations on the basis of the off-shell Kadanoff-Baym equations
\cite{Kadanoff1,Kadanoff2} for Green's functions in phase-space
representation (in first order gradient expansion beyond the
quasiparticle approximation). The approach consistently describes
the full evolution of a relativistic heavy-ion collision from the
initial hard scatterings and string formation through the dynamical
deconfinement phase transition to the strongly-interacting
quark-gluon plasma (sQGP) as well as hadronization and the
subsequent interactions in the expanding hadronic phase. In the
hadronic sector PHSD is equivalent to the Hadron-String-Dynamics
(HSD) transport approach \cite{CBRep98,Brat97} that has been used
for the description of $pA$ and $AA$ collisions from SIS to RHIC
energies in the past and the partonic dynamics is based on the
Dynamical QuasiParticle Model (DQPM) \cite{DQPM1,DQPM2}, which
describes QCD properties in terms of single-particle Green's
functions (in the sense of a two-particle irreducible (2 PI)
approach) and reproduces lattice QCD results -- including the
partonic equation of state -- in thermodynamic equilibrium.
The PHSD approach \cite{PHSD1,PHSD2,Bratkovskaya:2011wp}
has been applied to $pp$, $p+A$ and
$A+A$ collisions from $\sqrt{s_{NN}}$ = 5 GeV up to 2.76
TeV providing a consistent description of hadronic single-particle spectra,
collective flow coefficients of hadrons as well as dilepton
radiation from the partonic and hadronic phase. Since transport
coefficients are no inherent parameters of PHSD we may ask 
the question: what are the transport coefficients from PHSD in
equilibrium - in particular as a function of temperature - that are compatible
with the experimental observations? Here we summarize our findings
from Refs. \cite{box,xx2,xx3}.

\section{Shear viscosity coefficient: the Kubo formalism}
In this Section we concentrate on the extraction of the shear
viscosity from the 'infinite' parton-hadron matter employing the
Kubo formalism. We simulate the `infinite' matter within a cubic box
with periodic boundary conditions at various values for the quark
density (or chemical potential) and energy density. The size of the
box is fixed to $9^3$ fm$^3$. The initialization is done by
populating the box with light ($u,d$) and strange ($s$) quarks,
antiquarks and gluons. The system is initialized out of equilibrium
and approaches kinetic and chemical equilibrium during it's
evolution by PHSD. If the energy density in the system is below the
critical energy density ($\varepsilon_c \approx $ 0.5 GeV/fm$^3$),
the evolution proceeds through the dynamical phase transition (as
described in Ref. \cite{box}) and ends up in an ensemble
of interacting hadrons. For more details we refer the reader to Ref.~\cite{box}.

The Kubo formalism relates linear transport coefficients such as
heat conductivity, shear and bulk viscosities to non-equilibrium
correlations of the corresponding dissipative fluxes, and treats
dissipative fluxes as perturbations to local thermal equilibrium
\cite{Green,Kubo}. The Green-Kubo formula for the shear viscosity
$\eta$ is given by \cite{GKformula}:
\begin{equation}
\eta=\frac{1}{T}\int d^3r\int\limits_{0}^{\infty}
dt\bigl\langle\pi^{xy}({\bf 0},0)\pi^{xy}({\bf r},t)\bigr\rangle,
\end{equation}
where $T$ is the temperature of the system, $t$ refers to the time after
the system equilibrates, which is set at $t=0$, while $\langle...\rangle$
denotes the ensemble average in thermal equilibrium and $\pi^{xy}$
is the shear component (traceless part) of the energy momentum
tensor $\pi^{\mu\nu}$:
\begin{equation}
\pi^{xy}({\bf x},t)\equiv T^{xy}({\bf
x},t)=\int\frac{d^3p}{(2\pi)^3}\frac{p^{x}p^{y}}{E}f({\bf x} ,{\bf
p};t),
\end{equation}
where the mean-field $U_s$ enters in the energy $E=\sqrt{{\bf
p}^2+U_s}$.

\begin{figure}
\includegraphics[width=5.5cm]{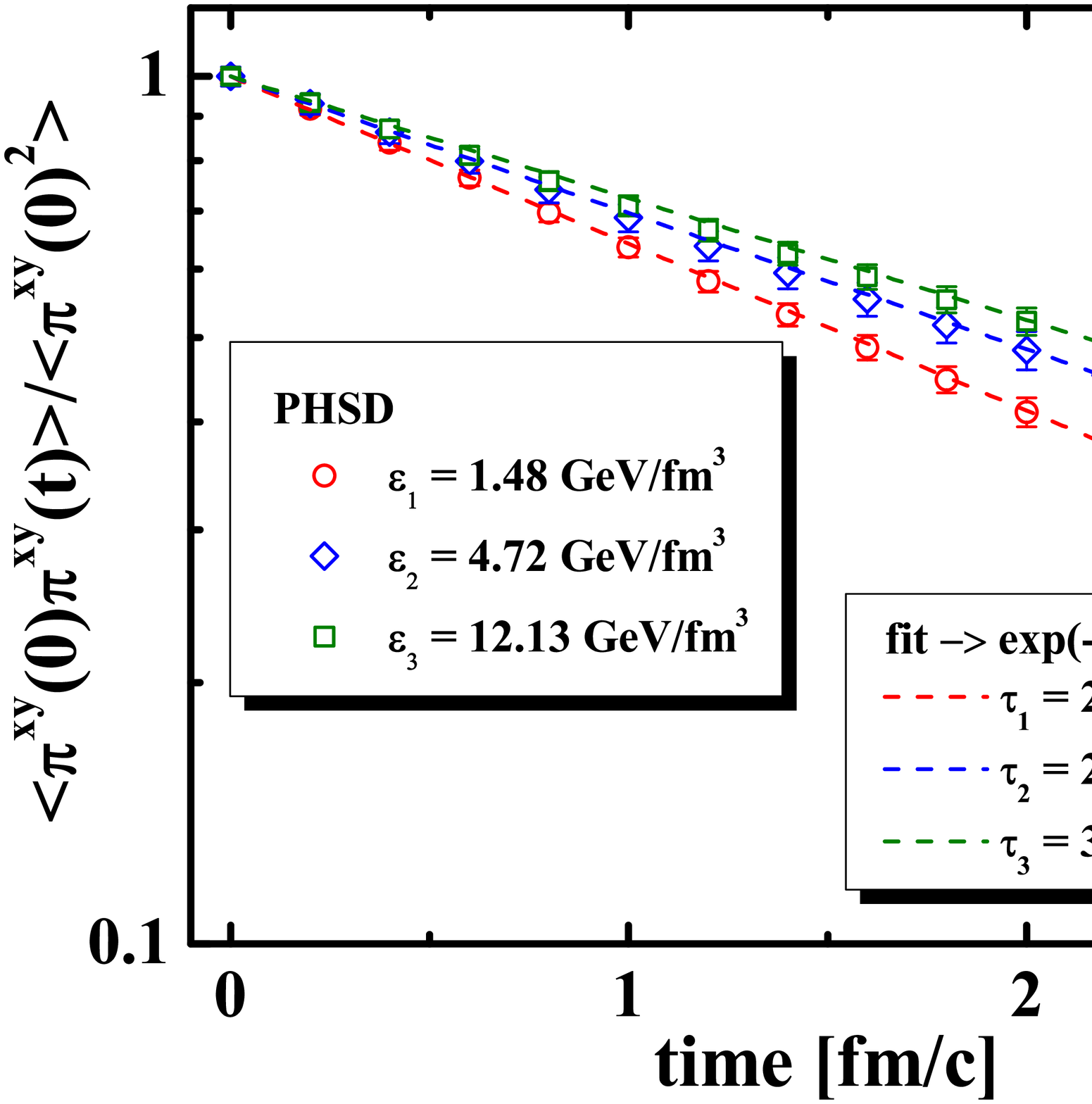}\hspace*{25mm}
\includegraphics[width=7.5cm]{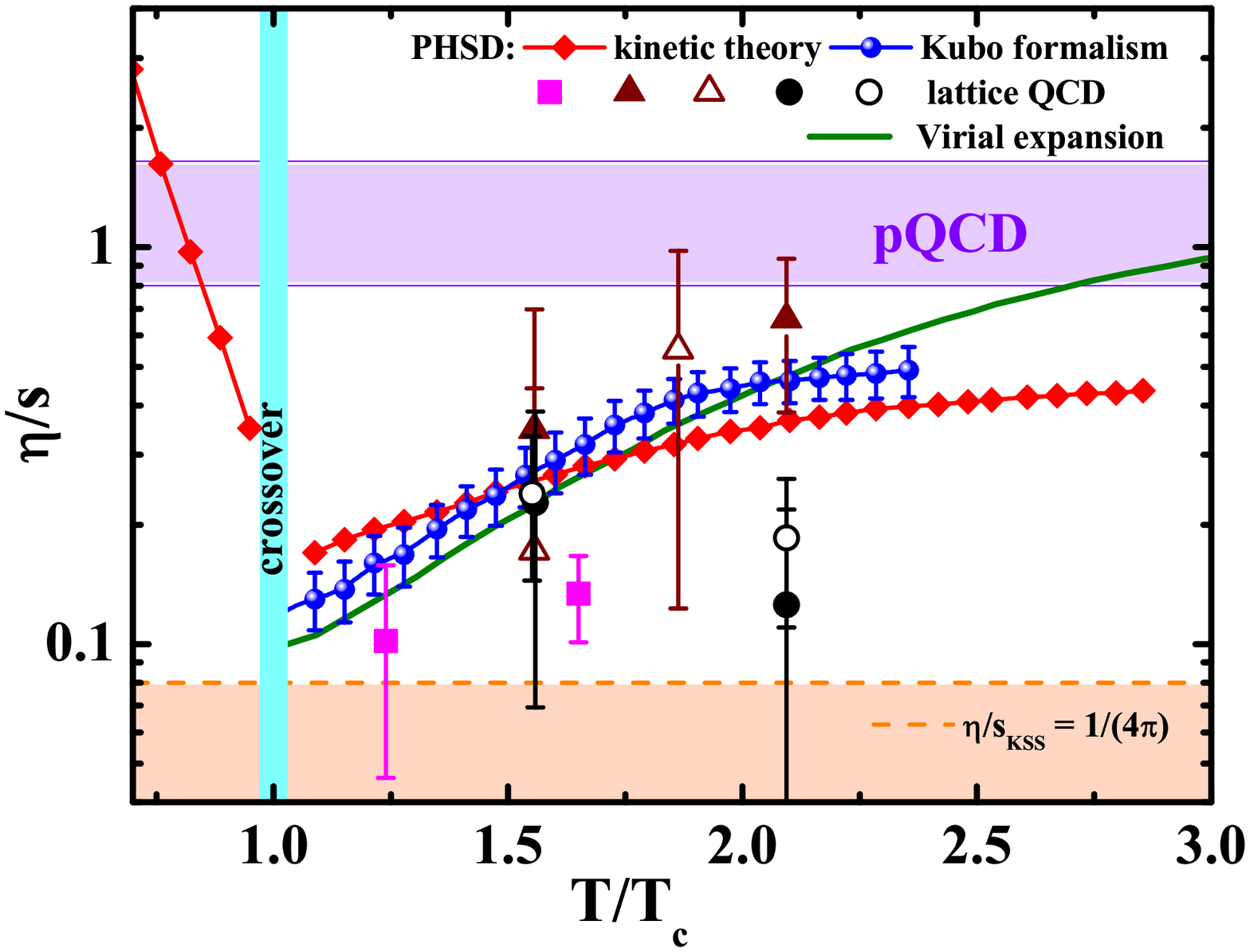}
\caption{ (lhs) The correlation functions
$\bigl\langle\pi^{xy}(0)\pi^{xy}(t)\bigr\rangle$, which are
normalized by $\bigl\langle\pi^{xy}(0)^2\bigr\rangle$, as a function
of time obtained by the PHSD simulations in the box (open symbols)
for systems at different energy densities and corresponding
exponential fits (dashed lines) with extracted relaxation
times.\\
(rhs) The shear viscosity to entropy density ratio
$\eta/s$ as a function of temperature of the system obtained by the
PHSD simulations using different methods: the relaxation time
approximation (red line$+$diamonds) and the Kubo formalism (blue
line$+$dots). The others symbols denote lattice QCD data for pure
$SU_c(3)$ gauge theory from \cite{lQCDtransport1} (magenta squares),
from \cite{lQCDtransport3} (wine open and full triangles), and from
\cite{lQCDtransport4} (black open and full circles). The orange
dashed line demonstrates the Kovtun-Son-Starinets bound \cite{KSS}
$(\eta/s)_{KSS}=1/(4\pi).$ For comparison, the virial expansion
approach (green line) \cite{Mattiello} is shown as a function of
temperature, too.}
\label{correlator}
\end{figure}

In our numerical simulation the volume averaged shear component of
the energy momentum tensor can be written as
\begin{equation}
\pi^{xy}(t)=\frac{1}{V}\sum\limits_{i=1}^{N}\frac{p_i^xp_i^y}{E_i},
\end{equation}
where $V$ is the volume of the system and the sum is over all particles
in the box at time $t$. Note that the scalar mean-field contribution
$U_s$ only enters via the energy $E$. Taking into account that point
particles are uniformly distributed in our box (implying
$\pi^{xy}({\bf r},t)=\pi^{xy}(t)$), we can simplify the Kubo formula
for the shear viscosity to
\begin{equation}
\eta=\frac{V}{T}\int\limits_{0}^{\infty}dt\bigl\langle\pi^{xy}(0)\pi^{xy}(t)\bigr\rangle.
\end{equation}
The correlation functions
$\bigl\langle\pi^{xy}(0)\pi^{xy}(t)\bigr\rangle$ are empirically
found to decay exponentially in time,
\begin{equation}
\bigl\langle\pi^{xy}(0)\pi^{xy}(t)\bigr\rangle=\bigl\langle\pi^{xy}(0)\pi^{xy}(0)\bigr\rangle\
e^{-t/\tau} \ ,
\end{equation}
as shown in Fig.~\ref{correlator} (lhs), where $\tau$ is the
relaxation time. Finally, we end up with the Green-Kubo formula for
the shear viscosity,
\begin{equation}
\eta=\frac{V}{T}\bigl\langle\pi^{xy}(0)^2\bigr\rangle\tau,
\end{equation}
which we use to extract the shear viscosity from the PHSD
simulations in the box.

\section{The relaxation time approximation}
\subsection{Calculation of the shear and bulk viscosities}
The starting hypothesis of the relaxation time approximation (RTA)
is that the collision integral can be approximated by
\begin{equation}
C[f]=-\frac{f-f^{eq}}{\tau},
\end{equation}
where $\tau$ is the relaxation time. In this approach it has been
shown that the shear and bulk viscosities (without mean-field or
potential effects) can be written as \cite{Hosoya,Gavin,Kapusta}:
\begin{equation}
\eta=\frac{1}{15T}\sum\limits_{a}\int\frac{d^3p}{(2\pi)^3}\frac{|{\bf
p}|^4}{E_a^2}\tau_a(E_a)f^{eq}_a(E_a/T),
\end{equation}
\begin{equation}
\zeta=\frac{1}{9T}\sum\limits_{a}\int\frac{d^3p}{(2\pi)^3}\frac{\tau_a(E_a)}{E_a^2}
\bigl[(1-3v_s^2)E_a^2-m_a^2\bigr]^2 \ f^{eq}_a(E_a/T),
\end{equation}
where the sum is over particles of different type $a$ (in our case,
$a=q,\bar q,g$). In the PHSD transport approach the relaxation time can
be expressed in the following way:
\begin{equation}
\tau_a=\Gamma^{-1}_a,
\end{equation}
where $\Gamma_a$ is the width of particles of type $a=q,\bar q,g$,
which is defined within the DQPM.  In our numerical
simulation the volume averaged shear and bulk viscosities assume the
following expressions:
\begin{equation}
\eta=\frac{1}{15TV}\sum\limits_{i=1}^{N}\frac{|{\bf
p}_i|^4}{E_i^2}\Gamma^{-1}_i,
\end{equation}
\begin{equation}
\zeta=\frac{1}{9TV}\sum\limits_{i=1}^{N}\frac{\Gamma^{-1}_i}{E_i^2}\bigl[(1-3v_s^2)E_i^2-m_i^2\bigr]^2,
\end{equation}
where the speed of sound $v_s=v_s(T)$ is taken from \cite{lQCD}  or the DQPM, alternatively.
Note that $v_s(T)$ from both
approaches is practically identical since it is governed by the
DQPM, which reproduces the lattice QCD results.

In Fig.~1 (rhs) we present the shear viscosity to entropy
density ratio as a function of temperature of the system extracted
from the PHSD simulations in the box employing different methods:
the relaxation time approximation (red line$+$diamonds) and the Kubo
formalism (blue line$+$dots). For comparison, the virial expansion
approach (green line) \cite{Mattiello} and lattice QCD data for pure
$SU_c(3)$ gauge theory are shown as a function of temperature, too.

\subsection{Mean-field or potential effects}
\begin{figure}
\includegraphics[width=8cm]{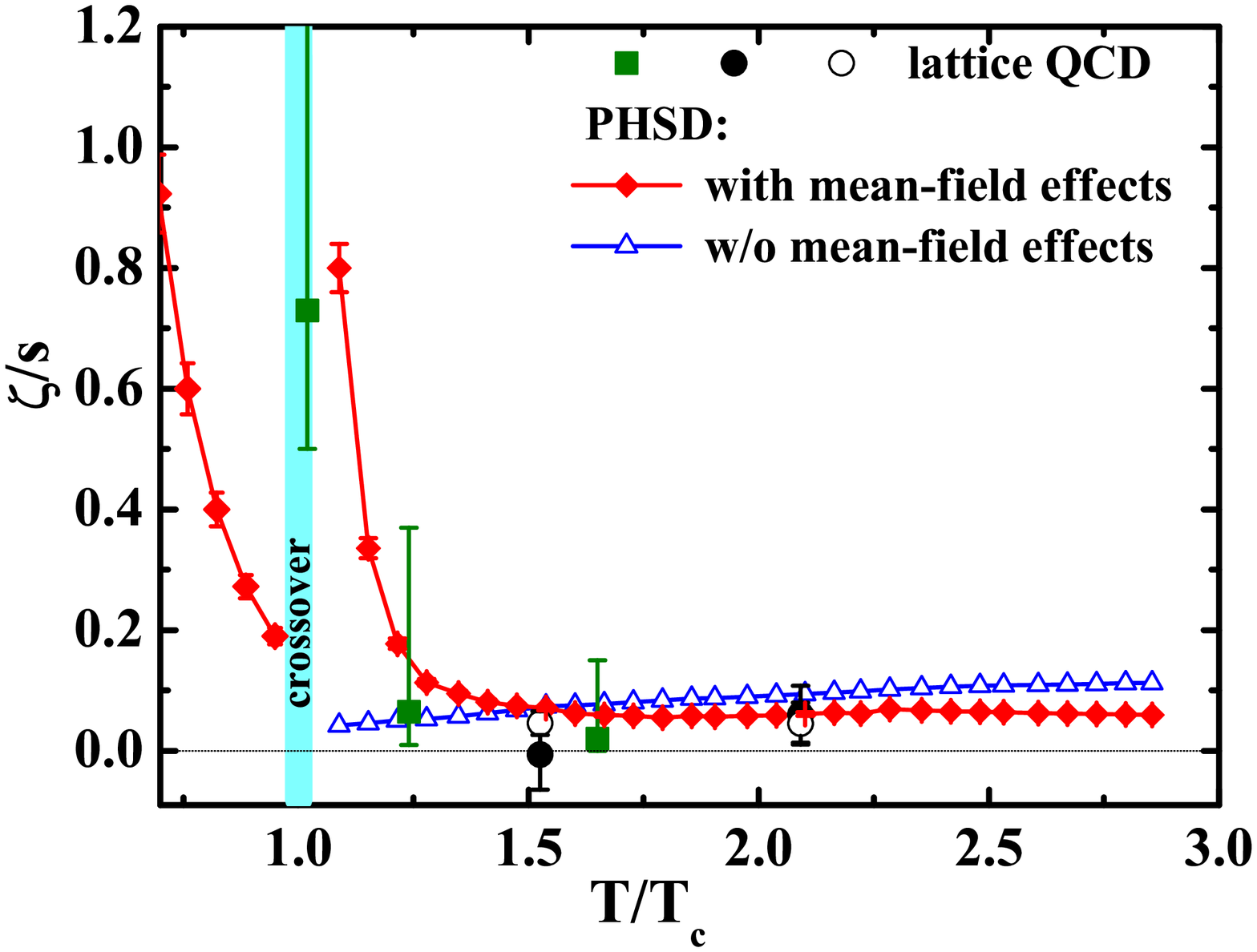}\includegraphics[width=8cm]{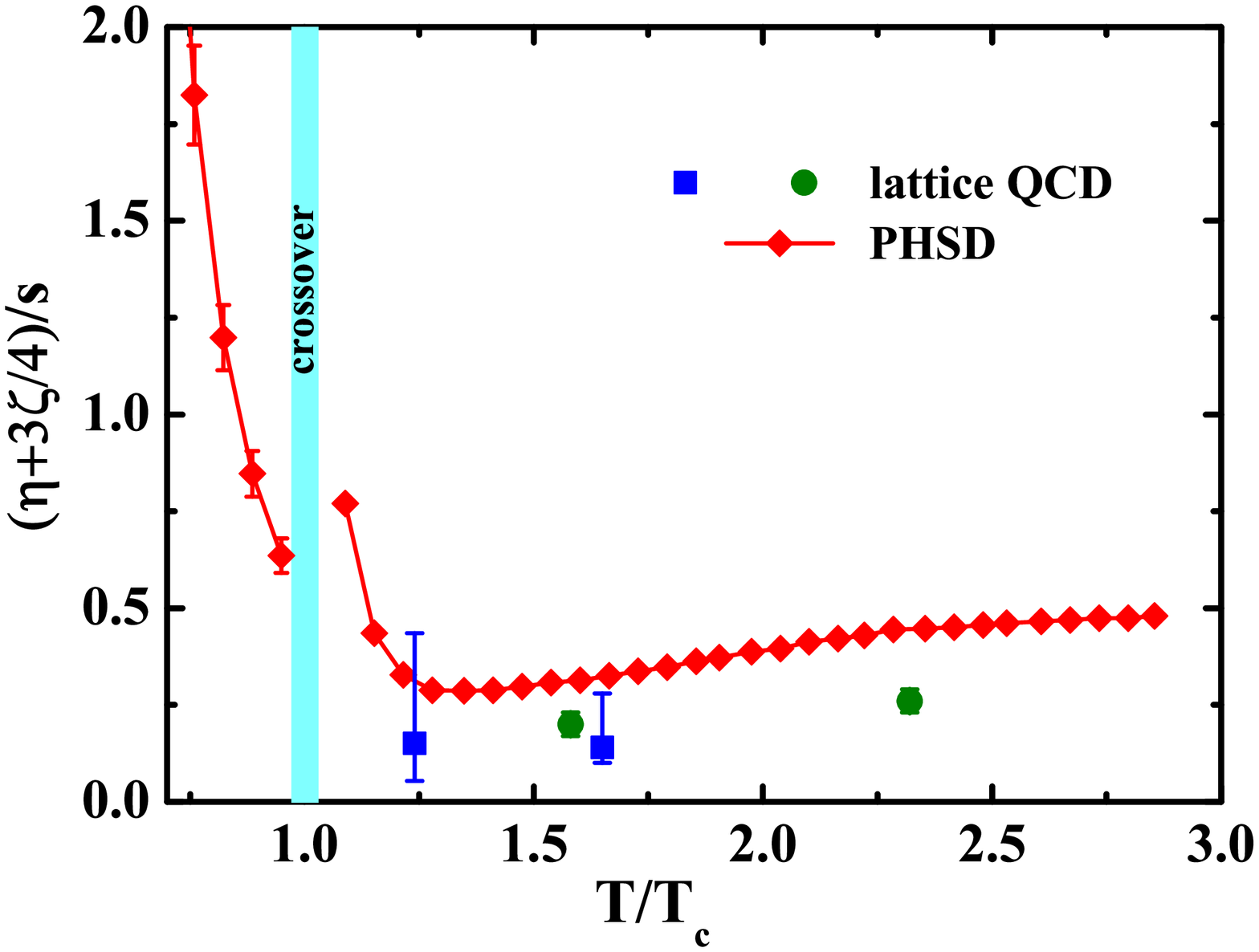}
\caption{ (lhs) The bulk viscosity to entropy density ratio
as a function of temperature of the system extracted from the PHSD
simulations in the box using the relaxation time approximation
including mean-field effects (red line$+$diamonds) and without
potential effects (blue line$+$open triangles). The available
lattice QCD data from \cite{lQCDtransport2} (green squares) and from
\cite{lQCDtransport4} (black open and full circles) are shown,
too. \\
(rhs) The specific sound channel
$(\eta+3\zeta/4)/s$ as a function of temperature of the system
obtained by the PHSD simulations in the box using the relaxation
time approximation including mean-field effects (red
line$+$diamonds). It is compared with lattice QCD data from
\cite{lQCDtransport5} (green circles) and from combining results of
\cite{lQCDtransport2} and \cite{lQCDtransport4} (blue
squares).}\label{bulk}
\end{figure}
In the absence of a chemical potential for quarks there should be no
sizeable vector or tensor fields, only scalar fields. This
affects the bulk viscosity, but not the shear viscosity. The
expression for the bulk viscosity with potential effects is given
by \cite{Kapusta}
\begin{eqnarray}
\nonumber\zeta
&=&\frac{1}{T}\sum\limits_{a}\int\frac{d^3p}{(2\pi)^3}\frac{\tau_a(E_a)}{E_a^2}f^{eq}_a(E_a/T)\\
&\times &\Bigl[\Bigl(\frac{1}{3}-v_s^2\Bigr)|{\bf
p}|^2-v_s^2\Bigl(m_a^2-T^2\frac{dm_a^2}{dT^2}\Bigr)\Bigr]^2.
\end{eqnarray}
In the numerical simulation the volume averaged bulk viscosity with
mean-field effects is used
\begin{equation}
\zeta=\frac{1}{TV}\sum\limits_{i=1}^{N}\frac{\Gamma^{-1}_i}{E_i^2}\Bigl[\Bigl(\frac{1}{3}-v_s^2\Bigr)|{\bf
p}|^2-v_s^2\Bigl(m_i^2-T^2\frac{dm_i^2}{dT^2}\Bigr)\Bigr]^2.
\end{equation}
Using the DQPM expressions for masses of quarks and gluons (for
$\mu_q=0$)
$$m_q^2(T)=\frac{1}{3}g^2(T) T^2,\,\,\,\,\,\,\,\,\,\,m_g^2(T)=\frac{3}{4}g^2(T) T^2$$
we can calculate the derivative $dm^2/dT^2$ as well
as $v_s^2$.

In Fig.~\ref{bulk} (lhs) we show the bulk viscosity to entropy density
ratio as a function of temperature of the system obtained by the
PHSD simulations in the box employing the relaxation time
approximation including mean-field (or potential) effects (red
line$+$diamonds) and without potential effects (blue line$+$open
triangles) as well as the available lattice QCD data from
\cite{lQCDtransport2,lQCDtransport4}.

A sound wave propagation in the $z$-direction with wavelength
$\lambda=2\pi/k$ is damped according to
\begin{equation}
T_{03}(t,k)\propto\exp{\Biggl[-\frac{\bigl(\frac{4}{3}\eta+\zeta\bigr)k^2t}{2(\varepsilon+p)}\Biggr]},
\end{equation}
where $T_{03}$ is the momentum density in the $z$-direction,
$\varepsilon$ is the energy density and $p$ is the pressure. Thus
both the shear $\eta$ and bulk $\zeta$ viscosities contribute to the
damping of sound waves in the medium.
In Fig.~\ref{bulk} (rhs) we present the specific sound channel
$(\eta+3\zeta/4)/s$ as a function of temperature of the system
obtained by the PHSD simulations in the box using the relaxation
time approximation including mean-field effects (red
line$+$diamonds).  It is compared with lQCD data from
\cite{lQCDtransport5} (green circles) and from combining results of
\cite{lQCDtransport2} and \cite{lQCDtransport4} (blue
squares).

\begin{figure}
\includegraphics[width=7.8cm]{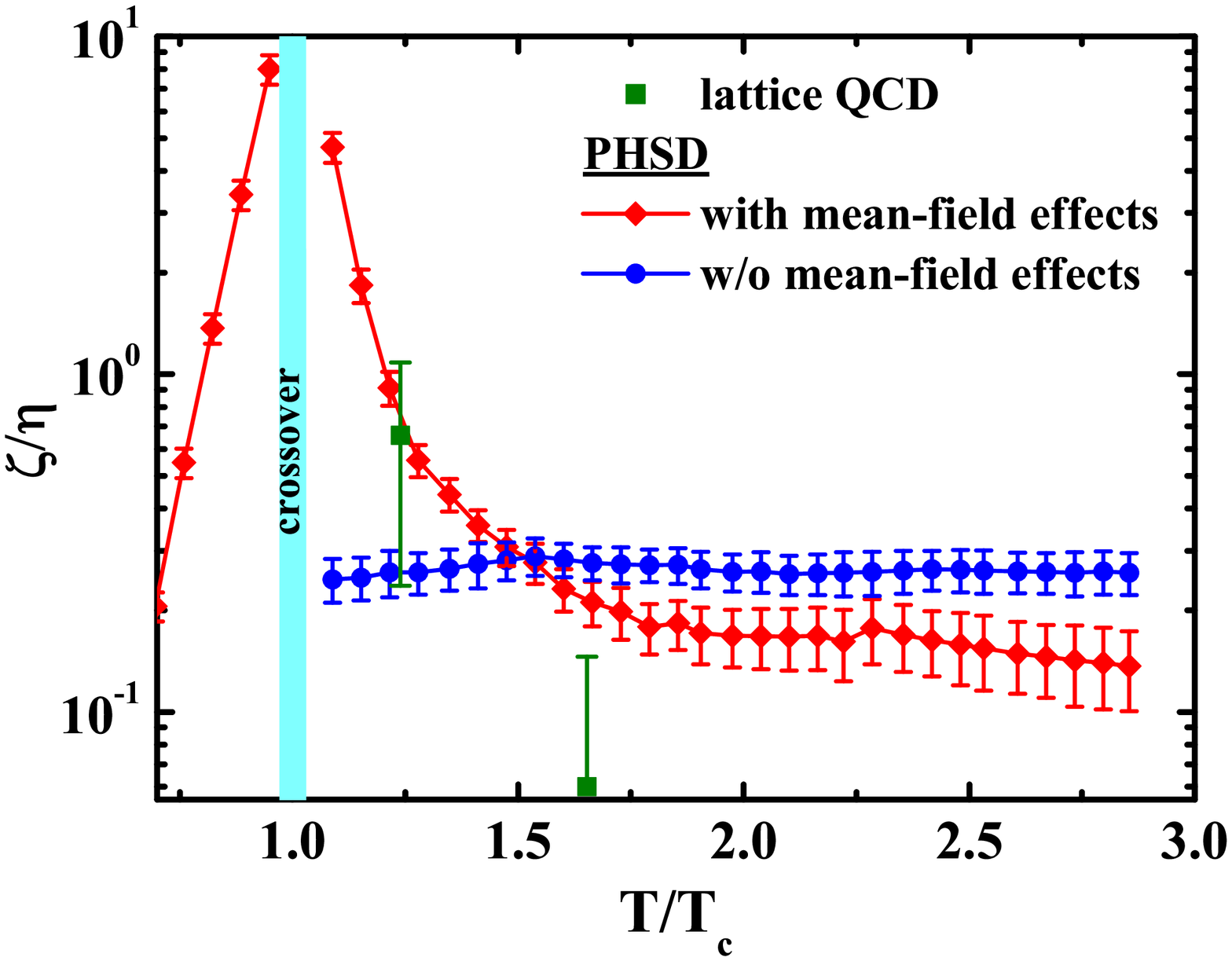}\includegraphics[width=8cm]{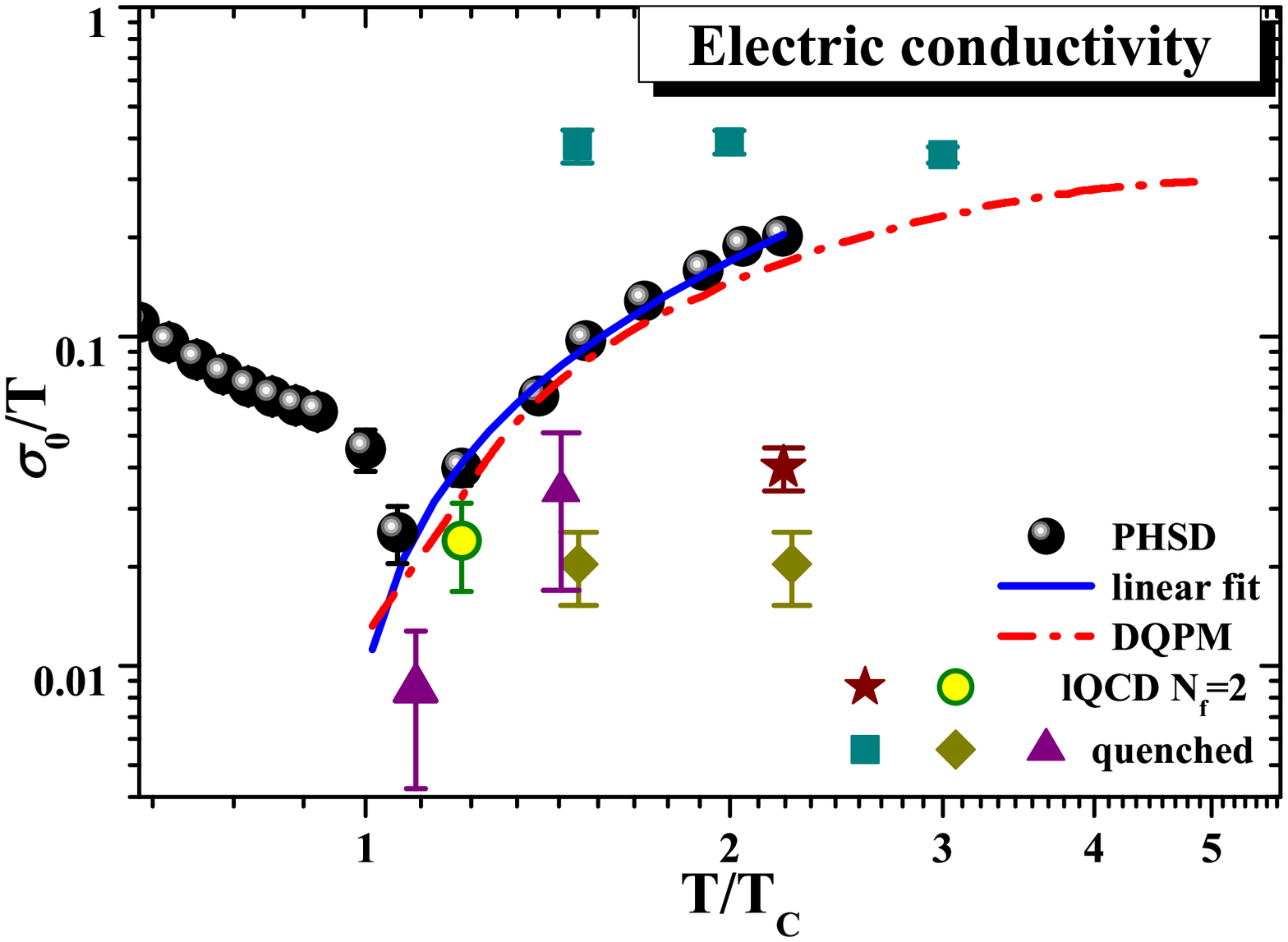}
\caption{ (lhs) The bulk to shear viscosity ratio $\zeta/\eta$ as a
function of temperature of the system obtained by the PHSD
simulations in the box employing the relaxation time approximation
including mean-field effects (red line$+$diamonds) and without
potential effects (blue line$+$circles). \\
(rhs) The ratio of the electric conductivity over temperature
$\sigma_0/T$ as a function of the
scaled temperature $T/T_c$ ($T_c$ = 158 MeV). The full round symbols
show the PHSD results, the solid blue line is a linear fit to the
PHSD results (above $T_c$), while the dash-dotted red line gives the
corresponding ratio in the relaxation-time approach (employing the
DQPM parameters). The scattered symbols with error bars represent
the results from lattice QCD calculations:  triangles --
Refs.~\cite{l1}, diamonds -- Ref.~\cite{l2},  squares --
Ref.~\cite{l3},  star -- Ref.~\cite{l4}, open circle --
Ref.~\cite{l5}.
}\label{ratio}
\end{figure}

Furthermore, in Fig.~\ref{ratio} (lhs), we show the bulk to shear viscosity
ratio $\zeta/\eta$ as a function of temperature of the system extracted from the
PHSD simulations in the box using the relaxation time approximation
including mean-field (or potential) effects (red line$+$diamonds)
and without potential effects (blue line$+$circles). This ratio
shows a pronounced maximum close to $T_c$ when including the
expected mean-field effects.

\subsection{Electric conductivity}
A further quantity of interest is the electric conductivity in the
partonic/hadronic matter which controls the electromagnetic
emissivity of the system in equilibrium. Here we briefly present
the results from Ref. \cite{xx3} which are displayed in Fig. 3
(rhs) for the dimensionless ratio of the electric conductivity over
temperature $\sigma_0/T$ which shows a pronounced minimum close to $T_c$ and
approaches a constant at higher temperature.

\section{Summary and conclusions}
We have employed the off-shell Parton-Hadron-String
Dynamics (PHSD) approach in a finite box with periodic
boundary conditions for the study of the shear and
bulk viscosities - as  well as the electric conductivity -
as a function of temperature (or energy
density) for dynamical infinite partonic and hadronic systems
in equilibrium. The PHSD transport model is based
on a lQCD equation of state [69] and well describes the
entropy density $s(T)$, the energy density $\varepsilon(T)$ as well
as the pressure $p(T)$ in thermodynamic equilibrium in
comparison to the lQCD results. We have
employed the Kubo formalism as well as the relaxationtime
approximation to calculate the shear viscosity $\eta(T)$.
We find that both methods provide very similar results
for the ratio $\eta/s$ with a minimum close to the critical
temperature $T_c$ while approaching the perturbative QCD
(pQCD) limit at higher temperatures. For $T < T_c$, i.e.
in the hadronic phase, the ratio $\eta/s$ rises fast with decreasing
temperature due to a lower interaction rate of
the hadronic system and a significantly smaller number
of degrees-of-freedom (or entropy density). Our results
are, furthermore, also in almost quantitative agreement
with the ratio $\eta(T)/s(T)$ from the virial expansion approach
in Ref. \cite{Mattiello} as well as with lQCD data for the
pure gauge sector.

We have, furthermore, evaluated the bulk viscosity
$\zeta(T)$ in the relaxation time approach and focused on the
effects of mean fields (or potentials) in the partonic phase.
Here we find a significant rise of the ratio $\zeta(T)/s(T)$ in
the vicinity of the critical temperature $T_c$ due to the
scalar mean-fields from PHSD. The result for this ratio
is in line with that from lQCD calculations. Additionally,
the specific sound channel ($\eta + 3\zeta/4)/s(T)$ has been calculated
and presents a non-trivial temperature dependence;
the absolute value for this combination of the shear and
bulk viscosities is in an approximate agreement with the
lattice gauge theory. Furthermore, the ratio $\zeta(T)/\eta(T)$
within the PHSD calculations shows a strong maximum
close to $T_c$, which has to be attributed to mean-field (or
potential) effects that in PHSD are encoded in the infrared
enhancement of the 'resummed' coupling $g(T)$ (from the DQPM).

We also find that the dimensionless ratio of the electric conductivity over temperature
$\sigma_0/T$ rises above $T_c$
approximately linearly with $T$ up to $T=2.5 T_c$, but approaches a
constant above $5 T_c$, as expected from pQCD. This finding is
naturally explained within the relaxation-time approach using the
DQPM spectral functions. Below $T_c$ the ratio $\sigma_0/T$ rises
with decreasing temperature because the system merges to a
moderately interacting gas of pions with a larger charge (squared) to mass
ratio than in the partonic phase and a longer relaxation time.

Since the PHSD calculations have proven to describe
single-particle as well as collective observables and also dilepton data  from relativistic
nucleus-nucleus collisions from lower SPS to top
RHIC energies \cite{ww1,ww2}, the extracted transport coefficients $\eta(T)$,
$\zeta(T)$ and $\sigma_0(T)$ are compatible with experimental observations
in a wide energy density (temperature) range. Furthermore, the
qualitative and partly quantitative agreement with lQCD
results is striking.

\vspace*{-4mm}
\section*{Acknowledgement}
 This work in part has been supported by
DFG as well as by the LOEWE center HIC for FAIR.

\end{document}